\documentclass[aps,prl,twocolumn,showpacs,a4paper,unsortedaddress,
amsmath,amssymb,byrevtex]{revtex4}

\usepackage[latin1]{inputenc}
\usepackage{graphicx}
\usepackage{dcolumn}
\usepackage{bm}
\usepackage{hyperref}
\usepackage{times}

\begin{document}
\preprint{ffuov/02-01}

\title{First-principles scheme for spectral adjustment
in nanoscale transport}

\author{V\'{\i}ctor M. Garc\'{\i}a-Su\'arez$^{1,2}$}
\author{Colin J. Lambert$^2$}

\affiliation{$^1$ Departamento de F\'{\i}sica, Universidad de
Oviedo \& CINN, 33007 Oviedo, Spain}\affiliation{$^2$ Department
of Physics, Lancaster University, Lancaster, LA1 4YB, United
Kingdom}

\date{\today}

\begin{abstract}
We implement a general method for correcting the low-bias
transport properties of nanoscale systems within an ab-initio
methodology based on linear combinations of atomic orbitals. We
show how the typical problem of an underestimated HOMO-LUMO gap
can be corrected, leading to quantitative and qualitative
agreement with experiments. We show that an alternative method
based on calculating the position of the relevant transport
resonances and fitting them to Lorentzians can significantly
underestimate the conductance and does not accurately reproduce
the electron transmission coefficient between resonances. We
compare this simple method in an ideal system of a benzene
molecule coupled to featureless leads to more sophisticated
approaches such as $GW$ and find a rather good agreement between
both. We also present results for a benzene-dithiolate molecule
between gold leads, where we study different coupling
configurations for straight and tilted molecules, and show that
this method yields the observed evolution of  two-dimensional
conductance histograms. We also explain the presence of low
conductance zones in such histograms by taking into account
different coupling configurations.
\end{abstract}


\maketitle

\section{Introduction}

Theories of coherent electron transport through nanostructures are
commonly based on a mean-field Hamiltonian $H$, describing a
scattering region connected to crystalline current-carrying leads.
Starting from $H$, a scattering matrix $S$ (or equivalent Greens
function) is calculated and transport properties such as the
electrical conductance $G$ obtained by evaluating Landauer-type
formulae. For example at zero temperature and bias,
$G=(2\mathrm{e}^2/\mathrm{h}) T(E_\mathrm{F})$, where $T(E)$ is
the transmission coefficient for electrons of energy $E$ passing
from one side of the scattering region to the other and
$E_\mathrm{F}$ is the Fermi energy. The mathematical machinery for
computing $S$ and $T(E)$ from $H$ is well established and is often
referred to as ``the scattering approach'' or ``non-equilibrium
Green's function approach'' to transport. When implemented with
the same level of rigor, these are mathematically equivalent. Such
theories have been used  to predict transport properties of a
large number of molecules between metallic leads
\cite{Tao06,Sel06,ash1,Lin07,Met08,Hai08,Fer09R,Kir09}.

The problem of identifying the most accurate mean-field
Hamiltonian for predicting nanoscale transport is a more open
question. Density functional theory (DFT) \cite{Hoh64,Koh65} is a
versatile tool for generating self-consistent, mean-field
Hamiltonians at both zero and finite applied voltages. For
metallic systems, such as atomic chains between electrodes of the
same material, there is quantitative agreement between experiment
and predictions derived from DFT-based mean-field Hamiltonians. In
contrast, for low-conductance systems such as single molecules
attached to metallic electrodes, predictions can differ from
experiments by orders of magnitude \cite{Lin07}, due to
limitations inherent to DFT \cite{Eve04,Sai05,Toh05,Koe06,Nea06}.
These discrepancies mainly arise from an
underestimation of the HOMO-LUMO (HL) gap, with theoretical
predictions of the order of 40\% of the experimental values
\cite{Per83,Sha83}. Approximated exchange-correlation functionals
also contain self-interaction errors \cite{Per81,Toh05}, which are
particularly important for localized states. Such errors, which worsen
the agreement between the removal energy and the last occupied
Kohn-Sham eigenvalue, produce an incorrect alignment between
the HOMO and the Fermi energy \cite{Toh07} and can also affect the width of the resonances when the coupling is small \cite{Ke07}. Even in
strongly-coupled molecules, such as  molecules with thiol anchor
groups, where the HOMO has a large weight on the sulphur atoms,
localized states can still have large self-interaction errors and
produce peaks very close to the Fermi energy. When the
self-interaction correction (SIC) is applied such states
move downwards and the zero-bias conductance decreases
\cite{Toh08}. Other approaches that avoid  the self-interaction
error, at least partially, are hybrid functionals, such as the
B3LYP \cite{Lee88,Bec93}, that include part of exact exchange
obtained from Hartee-Fock. Again, the main effect of such
functionals on the transport properties is the opening of the
HOMO-LUMO gap, specially when the molecule is strongly coupled to
the leads and charge transfer is small \cite{Ke07}. Exact exchange
approaches (EXX) such as Hartree-Fock or DFT with exact exchange
\cite{Yan02} also open the gap \cite{Ke07}. A complementary method to SIC is  LDA+$U$ \cite{Ani91,Coc05}, where a parameter $U$ is included to account for the intratomic repulsion, especially that produced on the $d$ states. The main effect of the $U$ is again the opening of the HOMO-LUMO gap, which reduces the zero-bias conductance \cite{Pem09}.  An alternative approach applies Hedin's $GW$ approximation on the extended molecule to correct its electronic properties \cite{Thy07,Dar07,Thy08-1,Thy08-2}. $ GW$ is
known to reproduce energy gaps more accurately and  also takes
into account screening effects of the leads. However, this method
has technical difficulties related to the connection between the
$GW$ corrected extended molecule and the leads and is also very
expensive. Consequently, to date, this method has only been applied
to small molecules.

Errors in the predicted spectrum of the DFT Hamiltonians lead to erroneous predictions for the positions of resonances in $T(E)$. To overcome this deficiency, spectral adjustment in nanoscale transport (SAINT) can be used to adjust the peak positions of $T(E)$
\cite{Mow08,Ceh08}. For a molecule attached to a metallic surface,
the correct positions of the HOMO and LUMO levels could be
determined experimentally from photoemission measurements.
Alternatively, they can be estimated theoretically by noting that
 the HOMO ($\varepsilon_\mathrm{H}$) and the LUMO
($\varepsilon_\mathrm{L}$) energies should coincide with the
negative of the ionization potential (IP) and electron affinity (EA), defined as $\mathrm{IP}=E_{N-1}-E_N$ and
$\mathrm{EA}=E_N-E_{N+1}$, where $N$ is the number of electrons in
the neutral molecule \cite{Cap97,Sau08}. These quantities can be
approximated by taking into account the Coulomb energy required to
charge the molecule, $E_\mathrm{C}$, as $\mathrm{IP}=
-\varepsilon_\mathrm{H}+E_\mathrm{C}$ and $\mathrm{EA}=
-\varepsilon_\mathrm{L}-E_\mathrm{C}$ \cite{CoulE}, so that the
charging energy is equal to $\mathrm{IP}-\mathrm{EA}=
\Delta_\mathrm{HL}+2E_\mathrm{C}$ \cite{Kaa08}, which is larger than the HL gap. However, when the molecule is placed in a polarizable environment such as metallic surfaces, the charging energy is reduced from the gas phase value by the polarization energies $P_+$ and $P_-$, which reduce both
the HL gap and the Coulomb energy. The change $\Delta\Sigma$ of
each frontier orbital can be decomposed into Coulomb-hole
($\Delta\Sigma_\mathrm{CH}$), screened exchange
($\Delta\Sigma_\mathrm{SX}$) and bare exchange or Fock
($\Delta\Sigma_\mathrm{X}$) contributions \cite{Nea06}. The bare
exchange is in general small and the screened exchange satisfies
$\Delta\Sigma_\mathrm{SX}\sim -2\Delta\Sigma_\mathrm{CH}$ for the
HOMO and $\Delta\Sigma_\mathrm{SX}\sim 0$ for the LUMO, so that
typically $P_+\sim -P_-$ \cite{Nea06}. From a physical point of view the reduction of the IP
can be understood by noting that the work required to take one
molecular electron to infinity is decreased, since the attraction
of the positively charged molecule that the electron leaves behind
is screened. The EA is increased because the positive image
charges help stabilize the extra electron placed on the molecule. Finally, to account for all effects, in the presence of coupling to the leads, a further shift occurs due the coupling self-energy.

\section{Theoretical approach}

Once the positions of the HOMO and LUMO resonances are known, the
question arises of how to obtain the most accurate
$T(E_\mathrm{F})$. In the literature there are two approaches to
this question and the aim of this paper is to compare the two. The
first method selects either the HOMO or LUMO resonance
$\varepsilon$ closest to the Fermi level and fits this to a
Lorentzian $T'(E) = \Gamma^2/[(E-\varepsilon)^2 + \Gamma^2]$,
whose width $\Gamma$ is obtained from the original DFT-based $H$.
The zero-bias conductance is then given by
$G=(2\mathrm{e}^2/\mathrm{h}) T'(E_\mathrm{F})$
\cite{Que07,Que09,Wan09}. This method has been shown to give
results in close agreement with experiments, but it is only
reliable when the relevant orbital is close to the Fermi level and
the transmission in the gap is described by the tail of a single
Lorentzian.

In our paper we compare this approach with an alternative method
for achieving SAINT, which adjusts the diagonal elements of the
mean-field Hamiltonian \cite{Mow08,Mar10} to reproduce known
values of the IP and EA. The starting point is a projection $\hat
H^0$ of the self-consistent mean-field Hamiltonian onto the atomic
orbitals associated with the molecule, whose eigenvalues and
eigenvectors are $\{\Psi_n\}_{n=1,...,N}$ and
$\{\epsilon_n\}_{n=1,...,N}$, respectively, where $N$ is the
number of atomic orbitals on the molecule. From this it is
possible to build a new Hamiltonian,

\begin{equation}
\hat H=\hat
H^0+\Delta_\mathrm{o}\sum_{n_\mathrm{o}}\left|\Psi_{n_\mathrm{o}}
\right>\left<\Psi_{n_\mathrm{o}}\right|+
\Delta_\mathrm{u}\sum_{n_\mathrm{u}}\left|\Psi_{n_\mathrm{u}}\right>
\left<\Psi_{n_\mathrm{u}}\right|
\end{equation}

\noindent whose eigenvalues corresponding to the occupied and
unoccupied levels are shifted by $\Delta_\mathrm{o}$ and
$\Delta_\mathrm{u}$, respectively. Taking into account the
definition of the density matrix, $\hat\rho=
\sum_{n=1}^N\left|\Psi_n\right>
\left<\Psi_n\right|f(\epsilon_n-\mu_\mathrm{e})$, where
$\mu_\mathrm{e}$ is the chemical potential, and the completeness
relation, $\hat 1= \sum_{n=1}^N\left|\Psi_n\right>
\left<\Psi_n\right|$, one notes that in the limit of zero
temperature $f(E-\mu_\mathrm{e})=\Theta(\mu_\mathrm{e}-E)$, and
 $\hat\rho=
\sum_{n_\mathrm{o}}\left|\Psi_{n_\mathrm{o}}\right>
\left<\Psi_{n_\mathrm{o}}\right|$, so that
$\sum_{n_\mathrm{u}}\left|\Psi_{n_\mathrm{u}}\right>
\left<\Psi_{n_\mathrm{u}}\right|=\hat 1-\hat\rho$. Finally,

\begin{equation}
\hat H=\hat H^0+(\Delta_\mathrm{o}-
\Delta_\mathrm{u})\hat\rho+\Delta_\mathrm{u}\hat 1
\end{equation}

\noindent In the general framework of non-orthogonal basis sets,
the matrix elements $H_{\mu\nu}$ of this Hamiltonian are obtained
by adding to the matrix elements of $\hat H^0$, the molecular matrix
elements of the density matrix operator and the identity operator
multiplied by the respective constants. In the case of the density
matrix one has

\begin{equation}
\left<\mu\right|\hat\rho\left|\nu\right>=
\sum_{n=1}^N\left<\mu\right|\left.\Psi_{n}\right>
\left<\Psi_{n}\right|\left.\nu\right>
f(\epsilon_n-\mu_\mathrm{e})=
\sum_{\alpha,\beta=1}^N S_{\mu\alpha}\tilde\rho_{\alpha\beta}S_{\beta\nu}
\end{equation}

\noindent where $\hat S$ is the overlap matrix. In this
expression, the matrix elements $\tilde\rho_{\alpha\beta}$ are defined to be $\tilde\rho_{\alpha\beta}=\sum_{n=1}^N
c_{n\alpha}c^*_{n\beta}f(\epsilon_n-\mu_\mathrm{e})$ where
\{$c_{n\alpha}$\} are coefficients in the expansion
$\left|\Psi_n\right>=
\sum_{\alpha=1}^Nc_{n\alpha}\left|\alpha\right>$.

The final result is a phenomenological Hamiltonian with
matrix elements

\begin{equation}
H_{\mu\nu}= H^0_{\mu\nu} + (\Delta_\mathrm{o}-
\Delta_\mathrm{u})\left<\mu\right|\hat\rho\left|\nu\right>+
\Delta_\mathrm{u} S_{\mu\nu}
\end{equation}

\noindent From this Hamiltonian one obtains a $T(E)$ which
possesses accurately-positioned resonances. In what follows, we
compare the zero-bias conductance $T(E_\mathrm{F})$ obtained from this
method, with the value obtained by making a Lorentzian fit to the
resonance closest to $E_\mathrm{F}$. We also compare this method to most sophisticated approaches such as the $GW$ approximation, for a benzene molecule between featureless leads, and study how the qualitative and quantitative trends in the conductance are modified when this method is used to correct the transmission of a benzene-dithiolate molecule between gold leads.

\section{Benzene ring between featureless leads}

To implement the above SAINT, we utilize the quantum transport
code SMEAGOL \cite{Roc06}, which uses the Hamiltonian provided by
the ab-initio code SIESTA \cite{Sol02} to compute
self-consistently the density matrix and the transmission
coefficients. To illustrate  how it works and compare it to other methods such as the $GW$ method we study first the simple case of a benzene
molecule coupled to featureless leads. We compare our results for the bias-dependent conductance to the results obtained by Thygesen and Rubio \cite{Thy08-1}.

In case of featureless leads, which act only as electron
reservoirs, the only effect of the coupling to the electrodes is a
broadening of the molecular levels by the imaginary part of the
self-energies, $\Gamma$. The retarded Green's function can
therefore be written as $\hat G^\mathrm{R}(E)=[(E+ 2i\Gamma)\hat S
- \hat H]^{-1}$. From the retarded Green's function we compute the
lesser Green's function, both in and out of equilibrium and obtain
the density matrix to make the process self-consistent
\cite{Roc06}. At the end of the self-consistent cycle, for a given
voltage, we apply the SAINT correction and calculate the density
of states (DOS), the transmission, which in this case can be
simplified to $T(E)=(i\Gamma/2)\mathrm{Tr}\{[\hat
G^\mathrm{R}(E)-\hat G^{\mathrm{R}\,\dag}(E)]\hat S\}$, and the
current. We define the SAINT shifts via the parameter $\bar\Delta=
(\Delta_\mathrm{o},\Delta_\mathrm{u})$.

We used a double-$\zeta$ polarized basis set (DZP) for both carbon and
hydrogen atoms and a real space grid defined with an energy cutoff of 200 Ry.
The DFT exchange and correlation energy was evaluated with the generalized
gradient approximation (GGA) as parametrized by Perdew, Burke and
Ernzerhof \cite{Per96}. With these approximations we obtain a HL gap of
5.99 eV and a quasiparticle gap, calculated by energy differences, of 10.00 eV, which is in very good agreement with previous results \cite{Thy08-1}.


We first analyze  how to implement the SAINT method. In the case
of a completely isolated molecule, the  method is unambiguous. However, when a molecule is
coupled to electrodes one has to separate the Hamiltonian and
density matrix of the molecule from those of the electrodes
\cite{MolDM}. The Hamiltonian of the whole extended molecule cannot be shifted, because that would introduce a discontinuity at the interface between the surfaces
of the extended molecule and the bulk leads. Therefore we project
onto orbitals belonging to atoms of the molecule and define $\hat H_0$ to be the sub-matrix of the Hamiltonian involving these orbitals only. The molecular density matrix is obtained by diagonalizing $\hat H_0$.

\begin{figure}
\includegraphics[width=\columnwidth]{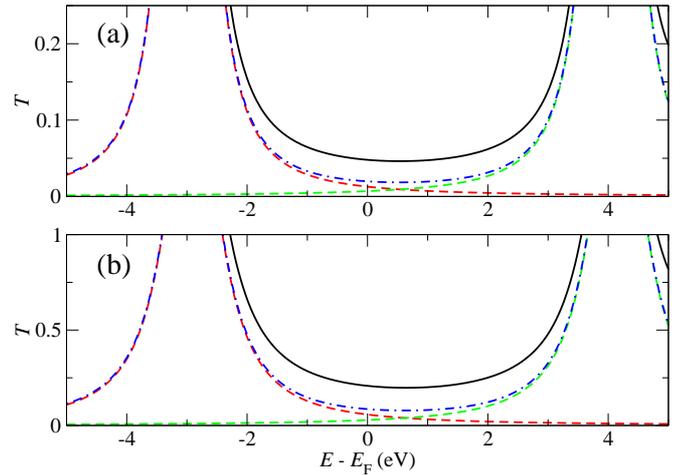}
\caption{\label{Fig2}Comparison between the full transmission curve
corrected with a SAINT $\bar\Delta=(-1,1)$ eV and the transmission obtained by fitting the HOMO or the LUMO to Lorentzians (dashed lines) and summing both (dashed-dotted line). $\Gamma=0.12$ eV in (a) and 0.25 eV in (b). Notice the different vertical scales on each panel.}
\end{figure}

\begin{figure}
\includegraphics[width=\columnwidth]{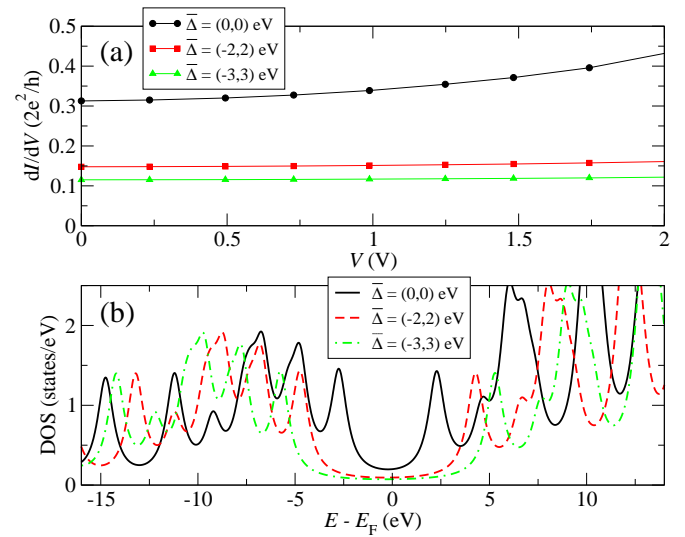}
\caption{\label{Fig3}Conductance obtained by numerical differentiation of the current (a) and the corresponding density of states at equilibrium (b) for various values of the
SAINT correction. $\Gamma=0.25$ eV.}
\end{figure}

In Fig. (\ref{Fig2}) we compare  results obtained by shifting the
molecular Hamiltonian levels with those obtained by fitting to
Lorentzians. Results are shown for the case $\bar\Delta=(-1,1) eV$
and two different couplings to the leads, to illustrate again the
effect of the broadening of the molecular levels. We fit
Lorentzians to the nearest resonances and compare these to the
$T(E)$ obtained from the spectrally-adjusted Hamiltonian. As can be
seen, simply by fitting to the nearest resonances, (either the
HOMO or the LUMO), the resulting zero-bias transmission is much
smaller than the SAINT result. Furthermore, adding both
Lorentzians does not significantly improve the results. This
disagreement arises because for a system with many resonances, the
Breit Wigner formula does not accurately describe the electron
transmission coefficient between resonances, even though it may be
accurate in the vicinity of a resonance \cite{oroz}.

We plot in Fig. (\ref{Fig3}) (a) the bias-dependent conductance
obtained by differentiating the current. The uncorrected DFT
result is very similar to the result of Thygesen and Rubio
\cite{Thy08-1}. It has however slightly larger value,  due to our
use of more basis states that produce additional resonances above
the Fermi level, in contrast with the minimal and truncated
Wannier function basis used in \cite{Thy08-1}, which gives rise to
only one resonance above the LUMO. We apply the SAINT correction
with $\bar\Delta=(-2,2)$ eV and $\bar\Delta=(-3,3)$ eV, which
roughly correspond to the correction to the HOMO and LUMO with and
without taking into account the correction due to the image
charges, respectively. The former would also approximately
correspond to the $GW$ case and the latter to the Hartree-Fock
case (HF). As can be seen, the SAINT corrected curves are
significantly lower and flatter than the DFT result, which agrees
with the $GW$ and HF results. This change in the conductance can
be explained by the opening of the gap and the subsequent reduction of the DOS in the middle of the gap, as can be seen in Fig. (\ref{Fig3})
(b), which reduces the transmission. Compared with \cite{Thy08-1},
the absolute values are again slightly higher, which may be due to
the use of different basis sets \cite{TrcBasis}. Since the main
effect of the $GW$ and HF methods is the opening of the gap, which
is exactly the same effect produced by SAINT, the origin of the
good agreement between these calculations is clear. The $GW$ and
HF give rise however to additional structure in the occupied and
unoccupied levels due to the energy-dependent $GW$ self-energy and
the different electronic structure method, respectively. Such
structure can slightly modify the value of the transmission in the
gap due to modifications in the tails of the resonances, but does
not  produce differences as large as those found between the exact
result and the Lorentzian approximation.

Two notes of caution should be added  when the SAINT method is
applied to more realistic systems. In some special cases, when the
Fermi level is pinned at the HOMO or the LUMO and the charge on
the molecule is significantly altered from its neutral value, the
self-consistent charge transfer, corresponding to that given by
the uncorrected DFT result, would introduce additional shifts in
the levels and possibly additional structure. Such charge transfer
would have been different if the gap had been bigger. Since the
SAINT is non-self-consistent and the gap is opened at the end, the
amount of charge transfer in such cases would be wrong. However the charge transfer in most of the systems is small \cite{Sta06} and
large distortions due to it are not expected to be common.
Another aspect that should be considered is related to molecular
states located on the coupling atoms and which penetrate into the
leads due to the strong coupling, such as the typical thiol-gold
connections \cite{Mow08}. In these systems such molecular states
have some weight inside the leads and therefore when the SAINT is
applied to only the molecular orbitals, these states can be
distorted. The main effect in the transport properties would be a change in the width of the associated resonances. These changes are not expected however to have an effect on the qualitative trends and would affect only slightly the quantitative predictions \cite{BDT-Au_s}. A special case, which would produce states that would penetrate deep into the leads, can occur when the coupling between the molecule and the leads is very strong and the on-site energies of both parts are very similar, but this is not very common in molecular electronics systems unless the molecule is very small \cite{Fer09T}.

\section{Benzene-ditholate between Au leads}

\begin{figure}
\includegraphics[scale=0.3]{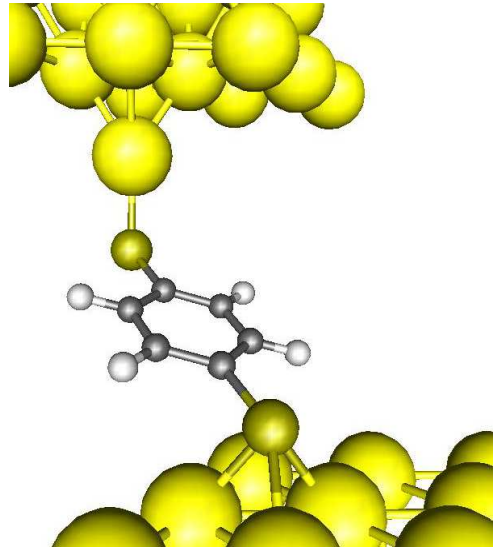}
\caption{\label{Fig4}A benzene-dithiolate molecule between (111) Au leads, tilted 70$^\circ$ from the normal and contacted on the bottom surface in a
hollow configuration and in the top surface to an adatom.}
\end{figure}

An archetypic example that has been extensively studied, both
experimentally \cite{Ree97,Xia04,Gho05,Ven06,Tsu06,Lor07,Mar08B}
and theoretically
\cite{Yal99,Ven00A,Ven00P,Xue01,Sto03,Bas03,Bra03,
Xue03,Tom04,Nar04,Ke05,Kon06,Gri06,Gar07,Li07} is the
benzene-dithiolate molecule coupled to gold leads. Recent experiments analyzed with 2D conductance histograms \cite{Mar08H,Que09,Kam09,Wan10}, on similar molecules (oligo(phenylene ethynylene) (OPE) molecules capped with thiols) contacted on one side to a surface and on the other side to a scanning tunneling microscope (STM), show a decrease of the
conductance as the separation between the tip and the surface
increases and a 2D zone with an upper and a lower edge of high and
low conductances, respectively. Some histograms also show at large
tip-surface separations a circular zone of low conductances
\cite{Que09,Wan10}. Such histograms suggest that these molecules
can be contacted with a tilted configuration for small tip-surface
separations; as the tip retracts from the surface, the angle
increases and the conductance is reduced. This can happen, as we shall see, if at least one of the sulphurs is contacted to an adatom, which
corresponds to the most probable coupling configuration to the
tip.

Many theoretical calculations have been carried out with the benzene
molecule oriented normal to the surfaces and coupled in a hollow configuration, which is predicted to be the most stable contact configuration and gives a strong coupling. When the molecule is contacted on top of a gold atom and normal to the surface, however, the coupling between the HOMO orbital, which is mainly made of perpendicular $x$ and $y$ $p$ orbitals, to the $s$ orbital of the gold adatom is rather small due to symmetry. When the molecule is tilted, the $x$ and $y$ orbitals of the sulphur increase the overlap with the adatom orbitals and the coupling increases. The same does not happen however when the sulphur is contacted in a hollow configuration
because the coupling of such orbitals to the gold surrounding
atoms is already large and it is not severely reduced by symmetry.
One would therefore expect a rather strong angular dependence when
one of the sulphurs is contacted to an adatom and a weaker angular
dependence when both sulphurs are contacted in a hollow
configuration. The additional low conductance zone in the histogram could be produced by another gold adatom on the surface, which would
increase the distance between the tip and the surface and further
decrease the conductance due to the small coupling on both sides.

\begin{figure}
\includegraphics[width=\columnwidth]{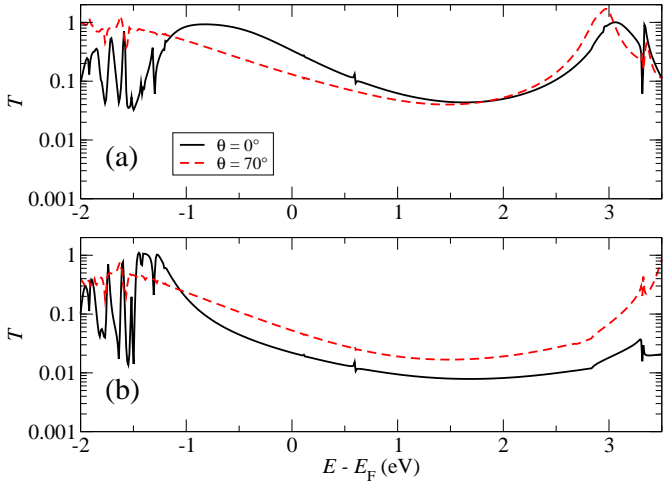}
\caption{\label{Fig5}Transmission of a benzene-dithiolate molecule
between (111) gold leads with both sulphur atoms contacted in the
hollow configuration and the molecule normally-aligned (continuous
line) or tilted (dashed line). The results were obtained without
(a) and with (b) the SAINT corrections given in table
(\ref{Tab01}).}
\end{figure}

In order to test these assumptions, we carried out ab-initio
calculations using the SMEAGOL code. The parameters were the same
as for the benzene molecule, but in this case we had to explicitly
include the leads. We chose (111) gold leads with 9 atoms per
slice, 2 slices on each side of the extended molecule, and three
additional slices to include the bulk leads. We also included
periodic boundary conditions along the perpendicular directions to
make sure the transmission coefficients were smooth. We chose a SZ
basis set for the gold atoms, which included the $s$ and $d$
orbitals \cite{BasisAu}. An example of one of the configurations,
where the molecule is tilted and contacted to an adatom in one of
the surfaces, can be seen in Fig. (\ref{Fig4}). We studied three
coupling configurations, hollow-hollow (HH), adatom-hollow (AH)
and adatom-adatom (AA), and for each case we compared the
normally-aligned molecule ($\theta=0^\circ$ from the normal to the
surface) to a tilted molecue ($\theta=70^\circ$).

\begin{figure}
\includegraphics[width=\columnwidth]{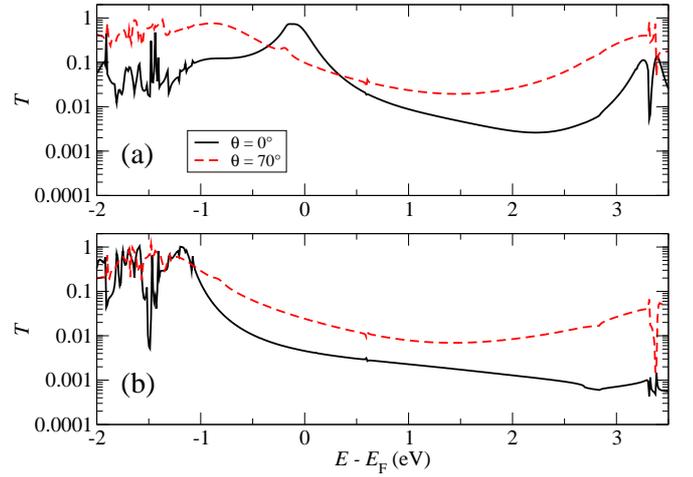}
\caption{\label{Fig6}Transmission of a benzene-dithiolate molecule
between (111) gold leads with one sulphur atom contacted in the
hollow configuration and the other on top of an adatom and the
molecule positioned normally-aligned (continuous line) or tilted
(dashed line). The results were obtained without (a) and with (b)
the SAINT corrections given in table (\ref{Tab01}).}
\end{figure}

Before showing the results it is necessary to comment a technical
detail. In case of BDT and other  molecules with degenerate
levels, one has to take additional care in order to obtain the
correct shifts. The HOMO resonance in the transmission
coefficients given by this molecule is composed of two resonances,
which correspond to two nearly degenerated molecular orbitals, one
of which is empty in the isolated molecule without leads \cite{Fer09T}. These states become occupied when the molecule is saturated with
hydrogens, which are added to the sulphurs, or when the molecule
is coupled to the leads. Consequently, if levels are shifted using
the occupation obtained from the Hamiltonian of the isolated
molecule, the empty level associated to the HOMO moves upwards,
which is not the desired outcome. In such cases, to avoid this problem, the LUMO has to be shifted downwards by $\Delta_\mathrm{o}$ and only the LUMO+1 and higher-energy eigenstates shifted upwards.

\begin{figure}
\includegraphics[width=\columnwidth]{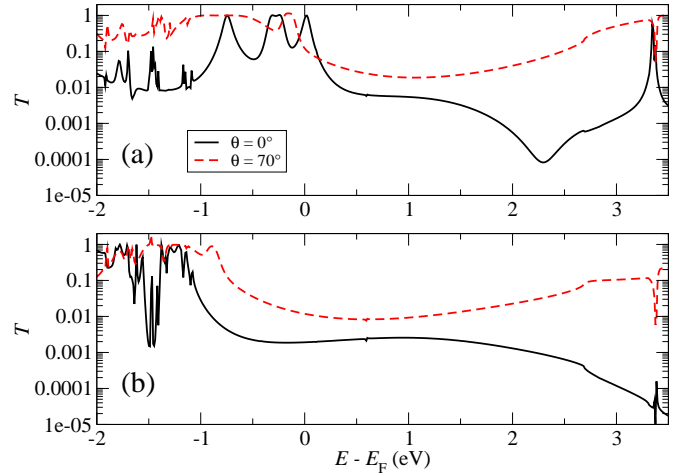}
\caption{\label{Fig7}Transmission of a benzene-dithiolate molecule
between (111) gold leads with both sulphur atoms contacted on top
of an adatom and the molecule positioned normally-aligned
(continuous line) or tilted (dashed line). The results were
obtained without (a) and with (b) the SAINT corrections given in
table (\ref{Tab01}).}
\end{figure}

In order to define the precise SAINT shifts it is necessary to know the DFT HOMO and LUMO, and the IP and EA (obtained by total energy differences) of the molecule in the gas phase, and the image charge corrections. The DFT HOMO, DFT LUMO, IP and EA are, respectively, -4.68, -1.38, 7.19 and -1.22 eV, from where we obtain gas phase corrections of -2.51 and 2.60 eV, for the occupied and unoccupied levels, respectively. The precise gap varies however from configuration to configuration, because it depends on the distance between the molecule and the image charge planes, which changes with the coupling configuration and the tilting angle of the molecule. We assume the image charge plane is 1 \AA\ above the surface and the point charge is in the middle of the molecule, so that the charging contribution can be approximated by $W=(\mathrm{e}^2\mathrm{ln}2)/(8\pi\epsilon_0a)$ \cite{Que09}, where $a=d/2-1$, and $d$ is the distance between surfaces. The values of $a$, $W$ and the final corrections to the occupied ($\Delta_\mathrm{o}$) and unoccupied ($\Delta_\mathrm{u}$) states are shown in table (\ref{Tab01}).

\begin{table}
\caption{Distance between the image plane, located 1 \AA\ above the surface, and the center of the molecule ($a$), image charge correction ($W$) and final corrections to the occupied ($\Delta_\mathrm{o}$) and unoccupied ($\Delta_\mathrm{o}$) levels.} \label{Tab01}
\begin{ruledtabular}
\begin{tabular}{lcccc}
&$a$ (\AA)&$W$ (eV)&$\Delta_\mathrm{o}$ (eV)&$\Delta_\mathrm{u}$ (eV)\\
\hline
HH 0$^\circ$&4.34&1.15&-1.36&1.45\\
HH 70$^\circ$&2.68&1.86&-0.65&0.74\\
AH 0$^\circ$&4.54&1.10&-1.41&1.50\\
AH 70$^\circ$&2.88&1.73&-0.78&0.87\\
AA 0$^\circ$&4.74&1.05&-1.46&1.55\\
AA 70$^\circ$&3.07&1.62&-0.89&0.98\\
\end{tabular}
\end{ruledtabular}
\end{table}

The results for the HH configuration are shown in Fig.
(\ref{Fig5}). As said before, the HOMO peak in this system is composed of two almost degenerated resonances coming from bonding and antibonding molecular orbitals generated by the states of the sulphur atoms near the Fermi level \cite{Fer09T}. This peak is very high in energy due to self-interaction errors \cite{Toh07} and produces very high conductances, compared to experiments. When the molecule is tilted, the peak widens and moves down in energy due to the increase of the coupling \cite{LUMO_benz}. Therefore, as can be seen in Fig. (\ref{Fig5}) (a),
for $\theta=70^\circ$, due to the movement of the HOMO to lower
energies, the zero-bias conductance of the normally-aligned
molecule turns out to be higher than that of the tilted molecule.
However, when the SAINT correction is applied, due to the facts that  the transmission of the normally-aligned molecule
decays faster than that of the tilted molecule and the
shift of the occupied levels in the first case is bigger than
the shift in the second case due to a smaller image charge
correction, when the SAINT corrections are applied the situation
reverses and the normally-aligned molecule gives a lower
conductance. The values of the transmission at the Fermi level can
be seen in table (\ref{Tab02}).

\begin{table}
\caption{Uncorrected and corrected zero-bias conductances $G$.} \label{Tab02}
\begin{ruledtabular}
\begin{tabular}{lcc}
&$G_\mathrm{uncorrected}$ (G$_0$)&$G_\mathrm{corrected}$ (G$_0$)\\
\hline
HH 0$^\circ$&3.3$\cdot$10$^{-1}$&2.2$\cdot$10$^{-2}$\\
HH 70$^\circ$&1.3$\cdot$10$^{-1}$&5.3$\cdot$10$^{-2}$\\
AH 0$^\circ$&5.0$\cdot$10$^{-1}$&4.5$\cdot$10$^{-3}$\\
AH 70$^\circ$&9.8$\cdot$10$^{-2}$&2.4$\cdot$10$^{-2}$\\
AA 0$^\circ$&8.1$\cdot$10$^{-1}$&1.9$\cdot$10$^{-3}$\\
AA 70$^\circ$&1.2$\cdot$10$^{-1}$&1.2$\cdot$10$^{-2}$\\
\end{tabular}
\end{ruledtabular}
\end{table}

In the normally-aligned AH configuration, shown in Fig.
(\ref{Fig6}) both the width and the height of the HOMO are smaller
than those in the HH configuration. The decrease of the width of
the resonance is due to the reduction of the coupling on the side
of the molecule coupled to the adatom and the decrease of the
height is due to the asymmetry of the contacts. When the molecule
is tilted, the width of the resonance increases dramatically and
again moves down in energy. The LUMO is also affected this time by
the increase of the coupling, which allows a better communication
between the states in the carbon rings and the states in the leads
than in the normally-aligned configuration. At zero bias and
without any correction, the conductance of the normally-aligned
molecule is again larger than that of the tilted molecule, which
does not agree with recent theoretical and experimental
predictions \cite{Wan10}. However, when the SAINT corrections are
applied, the faster decay of the transmission and the larger shift
of the occupied levels in the normally-aligned configuration again
reverse the trend and the tilted configuration gives  the largest
conductance, as can be seen in table (\ref{Tab02}).

Finally, in the AA configuration, shown in Fig. (\ref{Fig7}), the
HOMO peak splits into a number of sharp resonances, one of which
is pinned at the Fermi level. This movement of the HOMO towards
the Fermi level in the AA or top-top (on top of surface gold
atoms) configurations has been attributed to a reduction of the
charge transfer, which moves the states upwards and pins the HOMO
at the Fermi energy \cite{Bra03}. This movement increases the zero
bias conductance and, at the same time, decreases  the
transmission in the gap. When the molecule is tilted, the HOMO
resonances merge into a single peak, whose width is much larger
than that of the resonances of the normally-aligned molecule. This
peak moves downwards again and gives a lower transmission at the
Fermi level than that of the normally-aligned case. However, when the SAINT corrections are applied we see again that the situation reverses
and now the tilted molecule gives a higher conductance. The
results can be seen in table (\ref{Tab02}).

The conductances observed in previous experiments range from 10$^{-4}$ G$_0$ \cite{Ree97} to 0.1 G$_0$ \cite{Tsu06}. The values obtained by Reed {\em et al.} \cite{Ree97} are rather small and could be produced by specially stretched configurations. The rest of values are roughly within the range of conductances we obtain after applying the SAINT correction. The differences between the high and low conductance zones in different molecules observed in recent experiments range between approximately 1 order of magnitude in pyridines \cite{Que09} and 2 orders of magnitude in OPE's \cite{Wan10}. In our case, which is similar to that of the OPE's, we can see a difference of roughly 1 order of magnitude between the highest and lowest corrected values (excluding the HH configuration, which we predict would not be very probable in STM experiments). Probably the low conductance zone could have lower conductances arising from stretched AA configurations.

\section{Summary}

In summary, we have shown how to implement a spectral adjustment
to the bare DFT-based mean-field Hamiltonian, which corrects the
positions of transmission resonances in $T(E)$. We found that this
method can improve both the qualitative and quantitative agreements
with experiments and is an improvement over an alternative
approach based on Lorentzian fits.

For the case of a benzene molecule coupled to featureless leads, we tested the accuracy of using Lorentzian fits to correct the transport
properties of molecular junctions and found that such an
approximation can severely underestimate the transmission in some
cases. We also compared our method to previous $GW$ results. We
found a very good qualitative agreement, although the quantitative
agreement was worsened by the presence of Lorentzian tails coming
from unoccupied states. The agreement improved however when
smaller numbers of basis functions were used.

We also used the SAINT method in a realistic junction of a
benzene-dithiolate molecule between gold leads, in an attempt to
explain the evolution of the conductance as a function of length
in 2D histograms. We found that it was necessary to  shift the
molecular levels in order to obtain results that could explain the
observed evolution, i.e. results where the zero-bias conductance
of a normally-aligned molecule coupled to one or two adatoms was
smaller than the zero-bias conductance of a tilted molecule. We
also predicted the existence of a low conductance zone in the 2D
histograms of these molecules located more than one order of
magnitude below the main conductance zone.

Finally, we remark that the SAINT method can be used to correcting mean-field Hamiltonians even when the IP, the EA and other corrections are not known from first principles, as is the case for long molecules such as those reported in \cite{ash1}. To obtain a corrected Hamiltonian in these cases, the SAINT shifts $\Delta_\mathrm{o}$ and $\Delta_\mathrm{u}$ can be treated as free parameters to adjust the transmission coefficient to yield agreement with some chosen property, such as the current-voltage characteristic, the temperature dependence of the conductance or the dependence of the conductance on a gate voltage in, for example, an electrochemical environment. Once $\Delta_\mathrm{o}$ and $\Delta_\mathrm{u}$ are chosen to fit one of these measurements, the SAINT-corrected Hamiltonian can then be used to predict the others.

\begin{acknowledgments}
V.M.G.S. thanks the Spanish Ministerio de Ciencia e Innovaci\'on, the EPSRC
and the Marie Curie European ITNs FUNMOLS and NANOCTM for funding.
\end{acknowledgments}


\begin{thebibliography}{99}
\bibitem{CoulE}
In general the Coulomb energies that enter in the definition of IP
and EA are not the same.
\bibitem{MolDM}
In case of a molecule coupled to featureless leads we have only the orbitals of the molecule. In that sense defining the Hamiltonian of the molecular part has no ambiguity.
\bibitem{TrcBasis}
We did tests with single-$\zeta$ polarized (SZP), double-$\zeta$ (DZ), and single-$\zeta$ (SZ) basis sets and indeed found that the conductace decreased to lower values and the relative differences converged to the results of Thygesen and Rubio \cite{Thy08-1}. The zero-bias conductances (in units of G$_0$) obtained for each basis set for $\bar\Delta=(0,0)$, (-2,2) and (-3,3) eV, respectively, were the following. DZP: 0.31, 0.15, 0.11; SZP: 0.29, 0.13, 0.10; DZ: 0.26, 0.12, 0.09; SZ: 0.23, 0.10, 0.05.
\bibitem{BDT-Au_s}
Tests performed with a BDT molecule coupled to gold leads with only $s$ orbitals, which produce a smooth DOS and allow to clearly observe the effect of the SAINT corrections, showed indeed reductions of the resonances width. However such reductions were small and the resonances were not distorted.
\bibitem{BasisAu}
A SZ basis set with $s$ and $d$ orbitals for the gold atoms allowed to perform quick enough calulations without compromising the accuracy of the results. Test performed with only $s$ orbitals, however, showed a strong reduction of the conductance due to the orthogonality of the gold $s$ orbital and the sulphur $p_x$ and $p_y$ orbitals when one or two adatoms were contacted to the molecule. The inclusion of $d$ orbitals in gold is therefore necessary to produce reasonable results in such configurations.
\bibitem {LUMO_benz}
The LUMO is however not much affected because this orbital is located in the carbon ring and isnot so strongly influenced by changes in the coupling. Note that in our calculation, the values of $T(E)$ associated with this orbital are not very well defined because it is located near the gold band-edge. This is an artifact arising from the use of only a SZ basis set in the leads. However, since in equilibrium the Fermi level is very close to the HOMO (even after applying the SAINT corrections), the values of $T(E)$ around the LUMO are not important for our study.
\end{thebibliography}
\end{document}